\pdfoutput=1
\documentclass{article}
\usepackage{waspaa21,amsmath,graphicx,url,times}
\usepackage{color}

\usepackage{xcolor}
\usepackage[linesnumbered,ruled,vlined]{algorithm2e}
\usepackage{mathrsfs}
\usepackage{caption}

\title{Cross-Domain Semi-Supervised Audio Event Classification using Contrastive Regularization}

\name{Donmoon Lee$^{1, 2}$, Kyogu Lee$^{1, 3}$}
\address{$^1$ Music and Research Group, Department of Intelligence and Information, \\ Seoul National University,\\ 
         $^2$ Cochlear.ai, $^3$ Artificial Intelligence Institute, Seoul National University \\
          \{lunideal, kglee\}@snu.ac.kr\\       
}

\begin{document}

\ninept
\maketitle

\begin{sloppy}

\begin{abstract}
  In this study, we proposed a novel semi-supervised training method that uses unlabeled data with a class distribution that is completely different from the target data or data without a target label. To this end, we introduce a contrastive regularization that is designed to be target task-oriented and trained simultaneously. In addition, we propose an audio mixing based simple augmentation strategy that performed in batch samples. Experimental results validate that the proposed method successfully contributed to the performance improvement, and particularly showed that it has advantages in stable training and generalization.
  
\end{abstract}


\section{Introduction}
\label{sec:intro}

Sound contains a lot of information, but unfortunately, most of the audio-based studies have focused on speech. One of the barriers is the lack of a high-quality dataset. 
Since labeling audio is time-consuming and expensive, it is difficult to establish a large-scale labeled audio dataset.
Datasets such as AudioSet \cite{gemmeke2017audioset} acquire a large amount of data, but the quality of the labels is not guaranteed, which causes an additional problem, weakly supervised learning. 

Self-supervised learning is one approach to use datasets with such large scale and unreliable labels. This can be done in a label-free manner and the training is based on the assumption that the meaning of the data must remain in the presence of noise or transformations. 
It aims to learn general-purpose expressions that can be used anywhere without a target task and can be used for various downstream tasks through transfer learning or fine-tuning.
Especially, the recent contrastive learning-based approach is promising as it shows similar or superior performance to supervised training in the image domain \cite{chen2020simple_simclr, chen2020big_simclrv2}. In the case of the audio domain, various self-learning methods for speech have been proposed \cite{pascual2019learning_pase, schneider2019wav2vec}, and recently, it has been expanded to various attempts targeting general audio \cite{niizumi2021byol_audio, wang2021multi_simclr}. 
However, the downside is that self-supervised learning itself is so general that the performance of certain downstream tasks is not always guaranteed \cite{xiao2020should}.

Semi-supervised learning \cite{zhou2014semi} also utilizes unlabeled data.  
It differs from self-supervised learning in that there is a clear target task in the training procedure. 
In other words, the goal of semi-supervised learning is to perform the target task with the help of unlabeled data.
In general, most studies consider unlabeled data sets to contain a large number of target classes \cite{berthelot2019mixmatch, sohn2020fixmatch}. Therefore, the focus is on finding available data in unlabeled datasets. In the implementation, the performance validation is done using labels for parts of large datasets.
However, this assumption may not always be correct. Unlabeled data is literally unknown data, so there is no guarantee that the target data will be included. 

In this study, we propose a cross-domain semi-supervised learning method that can be used when unlabeled data have a different class distribution than labeled data. The proposed method combines the concepts of semi-supervised and self-supervised training. More specifically, the network is trained to perform target tasks while regularized by unlabeled data using a contrastive learning approach. One advantage of our approach is that any unlabeled data can be used. It is also useful as it is applied in the form of additional losses, so it can be applied to almost all common networks. To the best of our knowledge, this is the first study to perform neural network-based semi-supervised training using unlabeled data with different label distributions.

Our contributions are as follows: 
\begin{itemize}
    \item We propose a cross-domain semi-supervised training that can also be applied to data of completely different classes from the target data.
    \item We present a simple but efficient mixing strategy for applying contrastive learning to the audio domain, which is batch-split mixing.
\end{itemize}

\begin{figure}[t!]
 \centerline{
    \includegraphics[width=0.92\columnwidth]{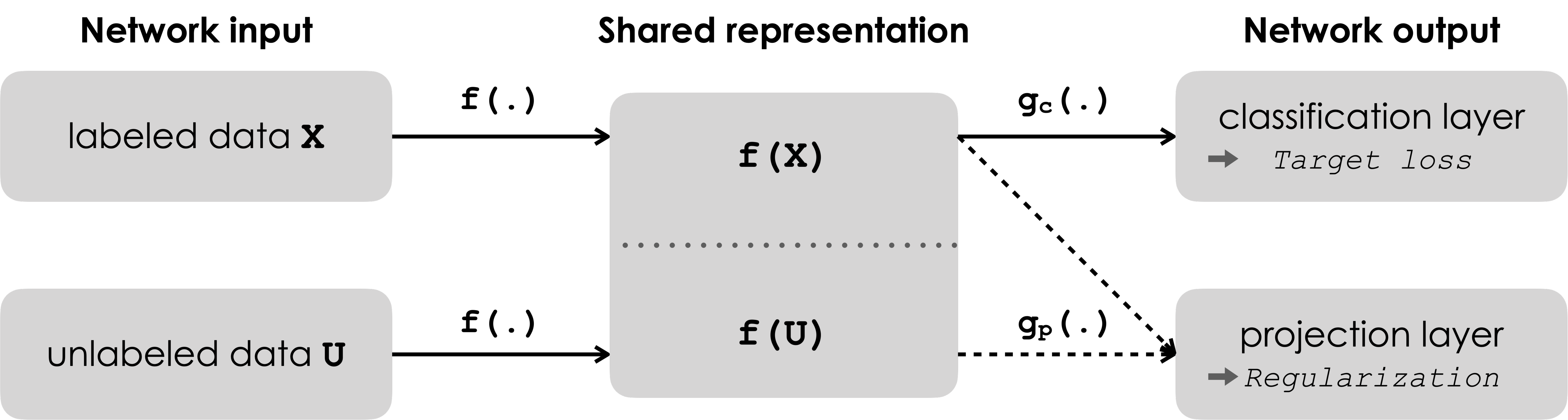}
    }
 \caption{The overview of the proposed cross-domain semi-supervised training. $f,  g_c, g_p$ are the shared feature extractor, classification layer, and L$^2$-normalized projection layer, respectively. The contrastive regularization is applied as an additional loss function.}
 \label{fig:overview}
\end{figure}

\section{contrastive regularization for semi-supervised learning}

\begin{figure}[t!]
 \centerline{
    \includegraphics[width=0.8\columnwidth]{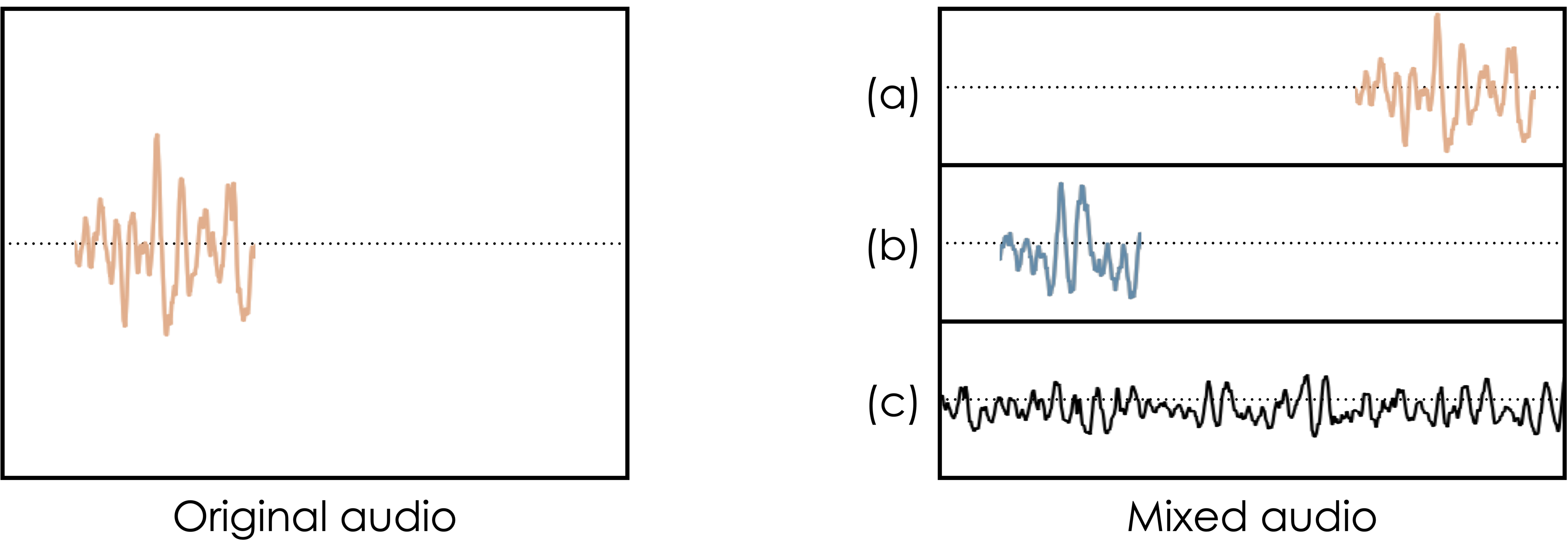}
    }
 \caption{The proposed audio augmentation strategy. The original audio sample is augmented with a) random offset, b) mixing with another source, and c) adding artificial noise.}
 \label{fig:mixing}
\end{figure}

The proposed semi-supervised method uses unlabeled data to learn differences between samples. 
As shown in Fig. \ref{fig:overview}, proposed regularization is performed in the form of multitask learning. 
However, it is different from the original multitask learning setup in that the regularization task uses a different unlabeled dataset than the target dataset. 
Rather, it is similar to the constraint method using other pre-trained networks \cite{johnson2016perceptual}, but the networks used for constraints are also trained at the same time. 
Such simultaneous training is expected to prevent common transfer learning problems such as catastrophic forgetting.

\subsection{Contrastive regularization}

Contrastive learning refers to a training method that recognizes identities and differences between each sample rather than classification through class labels \cite{bromley1993signature_siamese, hadsell2006dimensionality_siamese2, schroff2015facenet}.
This can also be applied to unlabeled data and is used to allow the network to learn semantics that is robust to data transformations. 
In particular, introducing \textit{NTXent} loss \cite{chen2020simple_simclr} to solve contrastive learning as an in-batch classification is a promising method in recent self-supervised studies.

\begin{equation}
    NTXent = - log \frac{exp(x \cdot x^{+}/\tau)}{exp(x \cdot x^{+}/\tau) + \sum_{x^{-}}exp(x \cdot x^{-}/\tau)}
\end{equation}

The proposed method introduces \textit{NTXent} loss in semi-supervised learning. 
It differs from the original method \cite{chen2020simple_simclr} in that the input of the \textit{NTXent} loss is one original sample and one augmented sample. 
We postulate that it will be more useful to the target task as it directly includes the target dataset as opposed to the original method of assuming a more general-purpose task.
This is a similar approach to the case where contrastive learning is used for a specific purpose.
In \cite{chang2021neural}, contrastive learning is applied to extract audio fingerprint. The learning process of contrastive learning can be viewed as a fingerprint task performed on a subset of the database. 
Therefore, in this case, the successful learning of contrastive learning guarantees the performance of the target task. 

%%% Coloring the comment as blue
\newcommand\mycommfont[1]{\footnotesize\ttfamily\textcolor{blue}{#1}}
\SetCommentSty{mycommfont}

\SetKwInput{KwInput}{Input}                % Set the Input
\SetKwInput{KwOutput}{Output}
\SetKwInput{KwNets}{Nets}

\begin{algorithm}[t]
\label{alg:pseudo_code}
\DontPrintSemicolon
  
  \KwInput{labeled dataset $(X,Y)$, unlabeled dataset $U$} 
  \KwNets{shared feature extractor $f(.)$, classificaton layer $g_c(.)$, and L$^{2}$ projection layer $g_p(.)$}

   \vspace{2pt}
   $f, g_{c} \gets$ WarmUp$(X,Y)$ \\ \tcp*{standard training for $f$ and $g_{c}$ }

  \While{epoch $<$ MaxEpoch}
    {
        \While{iter $<$ NumIters}
        {
            from $(X,Y)$, draw a mini-batch $\{x_b, y_b\}$ \\
            from $U$, draw a mini-batch $\{u_b\}$ \\
                 
            $s_1, s_2$ = Split(Shuffle(Concat($x_b, u_b$))) \\ \tcp*{all batch data divided in half }
            $\overline{s}_1 = $ Mixing($s_1, s_2, noise, SNR_{s1}$) \\
            $\overline{s}_2 = $ Mixing($s_2, s_1, noise, SNR_{s2}$) \\ \tcp*{augmented samples}

            $L_{clf} = CrossEntropy(y_b, g_{c}\cdot f(x_b))$\\ \tcp*{classification loss}       
            
            $L_{reg}^{(1)}$ =$ NTXentLoss(g_{p}\cdot f(s_1), g_{p}\cdot f(\overline{s}_1))$\\
            $L_{reg}^{(2)}$ =$ NTXentLoss(g_{p}\cdot f(s_2), g_{p}\cdot f(\overline{s}_2))$\\ \tcp*{regularization loss}
            $ Loss = L_{clf} +\lambda _{reg}  (L_{reg}^{(1)}+L_{reg}^{(2)})$ \\ \tcp*{total loss}
            $f, g_{c}, g_{p}\gets$ Optimize$(Loss)$ %\\ \tcp*{optimizing entire networks}
        }
    }

\caption{Training procedure of the cross-domain semi-supervised learning.}
\end{algorithm}

% \begin{algorithm}[t]
% \label{alg:pseudo_code}
% \DontPrintSemicolon
  
%   \KwInput{labeled dataset $(X,Y)$, unlabeled dataset $U$} 
%   \KwNets{shared feature extractor $f(.)$, classificaton layer $g_c(.)$, and L$^{2}$ projection layer $g_p(.)$}

%   \vspace{2pt}
%   $f, g_{c} \gets$ WarmUp$(X,Y)$ \\ \tcp*{standard training for $f$ and $g_{c}$ }

%   \While{epoch $<$ MaxEpoch}
%     {
%         \While{iter $<$ NumIters}
%         {
%             from $(X,Y)$, draw a mini-batch $\{x_b, y_b\}$ \\
%             from $U$, draw a mini-batch $\{u_b\}$ \\
                 
%             $w_1, w_2$ = Split(Concat($x_b, u_b$)) \\ \tcp*{all audio data divided in half }
%             $\overline{w}_1 = $ AudioMixing($w_1, w_2$) \\
%             $\overline{w}_2 = $ AudioMixing($w_2, w_1$) \\ \tcp*{augmented samples}

%             $L_{clf} = CrossEntropy(y_b, g_{c}\cdot f(x_b))$\\ \tcp*{classification loss}       
            
%             $L_{reg}^{(1)}$ =$ ContrastiveLoss(g_{p}\cdot f(w_1), g_{p}\cdot f(\overline{w}_1))$\\
%             $L_{reg}^{(2)}$ =$ ContrastiveLoss(g_{p}\cdot f(w_2), g_{p}\cdot f(\overline{w}_2))$\\ \tcp*{regularization loss}
%             $ Loss = L_{clf} +\lambda _{reg}  (L_{reg}^{(1)}+L_{reg}^{(2)})$ \\ \tcp*{total loss}
%             $f, g_{c}, g_{p}\gets$ Optimize$(Loss)$ \\ \tcp*{optimizing entire networks}
%         }
%     }
% \caption{Training procedure of the cross-domain semi-supervised learning.}
% \end{algorithm}

\subsection{Audio augmentation strategy}

We only used audio mixing as an augmentation method based on the unique properties of audio. 
One attribute of audio that differs from an image is that individual objects could be recognized even if multiple sources occur at the same time. Unlike an image where one masks the other when another source is added, the audio can be distinguished even if the two sources are mixed.
We thought that this property alone could make audio augmentation difficult enough and could be applied to any kind of audio data. 
Audio mixing is often used in contrastive learning \cite{niizumi2021byol_audio,wang2021multi_simclr}, but mainly to add noise or background. 
The same idea as ours have been applied to triplet-based learning \cite{jansen2018unsupervised}, but not to contrast learning, to the best of our knowledge. 

To do this in a batch, we split the batch in two and mix it with another split. 
The mixing process involves mixing with random offsets, random amplitudes, and additional artificial noise, as shown in Fig. \ref{fig:mixing}. 
Audio events can be shorter than the length of the network input, so after the audio mixing process, they can exist individually on the time axis rather than as a mixture at the same time.
This can be a relatively easy problem that only requires ensuring time-domain invariance, so additional noise is added to make it difficult. 
The reason for using artificial noise is that the traditional approach of using environmental sounds as noise, rather than the traditional target class \cite{thiemann2013diverse_DEMAND}, is limited in use by what the target sound is.

\begin{table*}[ht!]
\centering
% \captionsetup{justification=centering}

\caption{The results for semi-supervised training using the data with mismatched label on the \textit{ESC10}. Performance refers to the averaged test accuracy (\%) and standard deviation for predefined 5-folds. \textit{best} means the highest performance in the entire training epochs and \textit{last} means the averaged performance of the last 20 epochs. Note that $\dagger$ is evaluated on different dataset distributions.}
\label{table:cross-domain-semi-sup}

\begin{tabular}{c|c|ccccc}
\hline
\multicolumn{2}{c|}{Data fraction} & 10\%       & 25\%       & 50\%       & 75\%       & 100\%      \\ \hline
Supervised        & \textit{best}  & 59.0 ± 5.0 & 67.2 ± 2.2 & 75.0 ± 2.9 & 76.2 ± 3.7 & 81.2 ± 1.6 \\
                  & \textit{last}  & 53.0 ± 5.8 & 60.2 ± 2.4 & 65.9 ± 1.6 & 66.8 ± 5.0 & 72.3 ± 1.3 \\ \hline
Supervised        & \textit{best}  & 54.0 ± 4.8 & 76.2 ± 7.1 & 86.5 ± 1.5 & 89.5 ± 1.7 & 93.0 ± 3.6 \\
(ImageNet init.)  & \textit{last}  & 49.9 ± 4.8 & 73.0 ± 7.7 & 82.5 ± 1.7 & 84.9 ± 1.9 & 89.1 ± 3.6 \\ \hline
Proposed          & \textit{best}  & 65.2 ± 1.7 & 78.5 ± 3.7 & 90.0 ± 1.4 & 93.0 ± 1.3 & 95.2 ± 4.0 \\
                  & \textit{last}  & 61.8 ± 0.4 & 75.6 ± 3.2 & 85.4 ± 1.9 & 87.6 ± 3.2 & 92.4 ± 2.5 \\ \hline
Supervised \cite{cances2021comparison_escssl}         &    & 67.8 ± 4.0 & 82.5 ± 5.7 & 88.3 ± 3.0 & - & 91.7 ± 2.0 \\
Deep-Co-Training$\dagger$ \cite{cances2021comparison_escssl}  & \textit{}      & 75.7 ± 5.3 & 89.2 ± 4.0 & 91.7 ± 5.1 & -          & - \\ 

\hline
\end{tabular}
\end{table*}

\subsection{Training procedure}

The actual training procedure is listed in Algorithm 1. At the first stage, the target classification network, which includes a shared feature extractor and classification layer, is initialized through warm-up training. 
We found that training the network from scratch eventually yields similar performance, but takes more time to converge.
Therefore, warm-up training is performed with a relatively high learning rate to shorten the training time. 

In the semi-supervised learning stage, the labeled data is used for typical classification as well as warm-up training. The difference is that regularization that uses both labeled and unlabeled at the same time has been added. 
This results in \textit{NTXent} loss after batch split mixing, as mentioned earlier. The total loss is expressed as the sum of the classification loss and regularization loss.

\section{Experiments}

\subsection{Dataset}

\begin{itemize}
    \item \textit{ESC10} and \textit{ESC50}: The dataset for environmental sound classification \cite{piczak2015dataset} has a total of 2,000 audio samples for various audio events. \textit{ESC10} and \textit{ESC50} have 40 excerpts of 5 seconds for each class. It has the predefined 5-fold validation split. 
    \item \textit{US8K}: The urban sound 8K \cite{Salamon:UrbanSound:ACMMM:14} contains 8,732 labeled sound excerpts less than 4 seconds of 10 environmental sounds. \\
    The audio length varies from 50 milliseconds to 4 seconds and it has a predefined 10-fold validation split.
\end{itemize}

\subsection{Network Architecture}

A network takes 5-seconds of audio as input. It then converted to dB-scale log mel spectrogram, a common feature used in audio analysis. For the STFT, the nfft size was 1,024 and the hop size was 230, and the frequency information from 50Hz to 10,000Hz was converted to a 128-bin mel frequency. Thereafter, the mono channel mel spectrogram is stacked three times on the channel axis with the same value, to use the pre-trained weights \cite{palanisamy2020rethinking_inpt, guzhov2020esresnet_inpt}, resulting in the shape of (128, 480, 3). 

\textit{ResNet50} \cite{he2016identity_resnet}, which is often used in self-supervised studies \cite{wang2021multi_simclr, grill2020bootstrap_byol} with comparable performance in audio event classification, and is known to be effective for similar temporal unit analysis \cite{won2020evaluation}, was used as a feature extractor to output a 2,048-dimensional representation.
The classification task branch consists of a set of 128-dimensional fully connected layer, batch normalization, and \textit{ReLU} activation, followed by the final output layer. 
The contrastive regularization branch is similarly connected through 128-dimensional fully connected layer, batch normalization, and \textit{ReLU} activation, and as an output, an L2-normalized 64-dimensional embedding vector. 
The final output of this branch means an embedding for each data, comparisons with another embedding can be done through a dot product, the larger it means the two data are more similar.

\subsection{Experiment configuration}
Two experiments were conducted with data in and out of the dataset. 

\begin{itemize}
    \item \texttt{Semi-supervised training using the data with mismatched label}: \textit{ESC10} was used as labeled data. We defined \textit{ESC40} as \textit{ESC50} excluding \textit{ESC10} and used as unlabeled data in the proposed method. As with self-supervised studies \cite{chen2020simple_simclr}, all unlabeled labels were always used, and parts of labels were used. \\
    Deep-Co-Training$\dagger$ \cite{cances2021comparison_escssl} is a method of pseudo-labeling unlabeled data using two networks that are trained to provide different predictive values and complement each other. A portion of \textit{ESC10} is used as labeled and the rest is used as unlabeled.
    
    \item \texttt{Cross-dataset semi-supervised training}: Both \textit{ESC50} and \textit{US8K} were used in the experiment, one for labeled data and the other for unlabeled data. That is, when classifying \textit{ESC50}, \textit{US8K} was used for contrastive regularization and vice versa.
\end{itemize}

To reduce the influence of other variables in the experiment, no changes such as augmentation were made to the data pipe in target classification. The networks are initialized with the weights learned from the ImageNet. This is because the kernel trained to analyze the image is also known effectively for the classification of environmental sounds \cite{palanisamy2020rethinking_inpt, guzhov2020esresnet_inpt}.

\subsection{Implementation details}
All audio data is resampled to 22,500 Hz and the scale is normalized based on energy. When each audio is used for network input, the short audio is zero-padded for up to 5 seconds.
Audio mixing was done in the SNR range of [-5, 20] dB.
A total of 6 artificial noises including blue, brown, grey, pink, violet, and white were used and added to less than 6 dB.

Warm-up was performed until the training accuracy was high enough, \textit{ESC10} and \textit{ESC50} performed 20 epochs, and \textit{US8K} performed 10 epochs.
\textit{Adam} optimizer \cite{kingma2014adam} with learning rate 1e-4 is used for warm-up training and with learning rate 1e-5 used for semi-supervised training. For the proposed method, we trained the network 100 epochs with 8 labeled and 32 unlabeled samples, and for supervised learning, 200 epochs were used because the test accuracy was relatively slow to converge. 

Weight for regularization loss ($\lambda_{reg}$) is set to 0.05 taking into account the scale of the loss value. 
The temperature of the regularization loss is set to 0.01. We experimentally confirmed that training was successful in the range of [0.001, 0.1].
All experiments were implemented using \textit{Tensorflow} \cite{tensorflow2015-whitepaper}, and the network structure and weights of \texttt{ResNet50V2} included in the library were used. 

\section{Results and Discussion}

\subsection{Cross-domain semi-supervised learning on \textit{ESC10}}

\begin{table}[t!]
\centering
\caption{Experimental results for cross-dataset semi-supervised training. Performance refers to the average accuracy and standard deviation for predefined k-fold validation.}
\label{table:cross-dataset}

\begin{tabular}{ccc}
\hline
Target data    & Unlabeld data  & Accuracy (\%)                       \\ \hline
\textit{ESC10} & \textit{-}     & 89.1 ± 3.6\\
\textit{ESC10} & \textit{ESC40} & 92.4 ± 2.5 \\
\textit{ESC50} & \textit{-}     & 73.6 ± 1.0\\
\textit{ESC50} & \textit{US8K}  & 83.5 ± 3.0 \\
\textit{US8K}  & \textit{-}     & 78.6 ± 4.3\\
\textit{US8K}  & \textit{ESC50} & 79.7 ± 4.3\\ 
\hline
\end{tabular}
\end{table}

% \begin{table}[ht]
% \centering
% \caption{Cross-dataset semi-supervised experiments. Performance refers to the average accuracy (\%) and standard deviation for predefined folds.}
% \label{table:cross-dataset}

% \begin{tabular}{ccc}
% \hline
% Target data    & Unlabeld data  & Accuracy (\%)                       \\ \hline
% \textit{ESC10} & \textit{-}     & 72.3 ± 1.3 (89.1 ± 3.6)\\
% \textit{ESC10} & \textit{ESC40} & 92.4 ± 2.5 \\
% \textit{ESC50} & \textit{-}     & 57.4 ± 1.3 (73.6 ± 1.0)\\
% \textit{ESC50} & \textit{US8K}  & 83.5 ± 3.0 \\
% \textit{US8K}  & \textit{-}     & 66.5 ± 5.2 (78.6 ± 4.3)\\
% \textit{US8K}  & \textit{ESC50} & 79.7 ± 4.3\\ 
% \hline
% \end{tabular}
% \end{table}

The results for semi-supervised training using the data with mismatched labels are listed in Table \ref{table:cross-domain-semi-sup}. \textit{best} is the average of the highest accuracy observed within the entire epochs at each fold and can be seen as the maximum performance that can be achieved in this experimental setup. \textit{last} is the average accuracy of the last 20 epochs and is generally expected to be achieved. 
Except for 10\%, an almost constant performance difference is observed between supervised learning using different initialization methods. 
Initializing the weights from the image domain had a huge impact on the performance improvement, reconfirming that the kernel trained in the image domain still helped the spectrogram domain.
However, the reversed performance at 10\% shows that these weighted initialization methods may harm the network when data is scarce. In this case, since the data is very small, 3 samples per class, it is thought that the network focuses on other elements learned from images rather than semantic elements, but more research is needed.

The proposed method shows the highest accuracy in all experimental settings.
Since the proposed method does not affect the classification layer structurally, it can be seen that a better semantic representation is induced by adding contrastive regularization. 
Multitask learning can improve target performance depending on the composition of the task \cite{caruana1997multitask, lee2019enhancing}, 
and in this case, it can be seen that it behaved as we intended.
% and in the proposed method it seems to enhance the representation in a way that improves the target task through multitask learning.
It also shows that even if each branch uses data from a different label distribution, or a completely different dataset (Table \ref{table:cross-dataset}), the multitask learning assumptions still work. The performance improvement of \textit{US8K} is less than that of \textit{ESC50}. Unlike \textit{ESC50}, \textit{US8K} consists of audio clips shorter than 5 seconds, and we assume that energy-based mixing is expected to adversely affect.

The proposed method has relatively low performance and performance improvement compared to regular semi-supervised training \cite{cances2021comparison_escssl} that uses the rest of labeled data as unlabeled data. 
Excluding the 10\% condition, the standard semi-supervised training shows a higher performance improvement and converges on performance with the full data at about 50\%. 
However, at 100\%, the proposed method showed higher performance even though there are no various performance improvement techniques.
This shows the difference between the standard method and the proposed method. The standard method always uses the same class of data under strong assumptions, so it can reach the desired performance faster with less data, but there is a limit to the performance gain.

\subsection{Generalization effect}

\begin{figure}[t!]
 \centerline{
    \includegraphics[width=0.87\columnwidth]{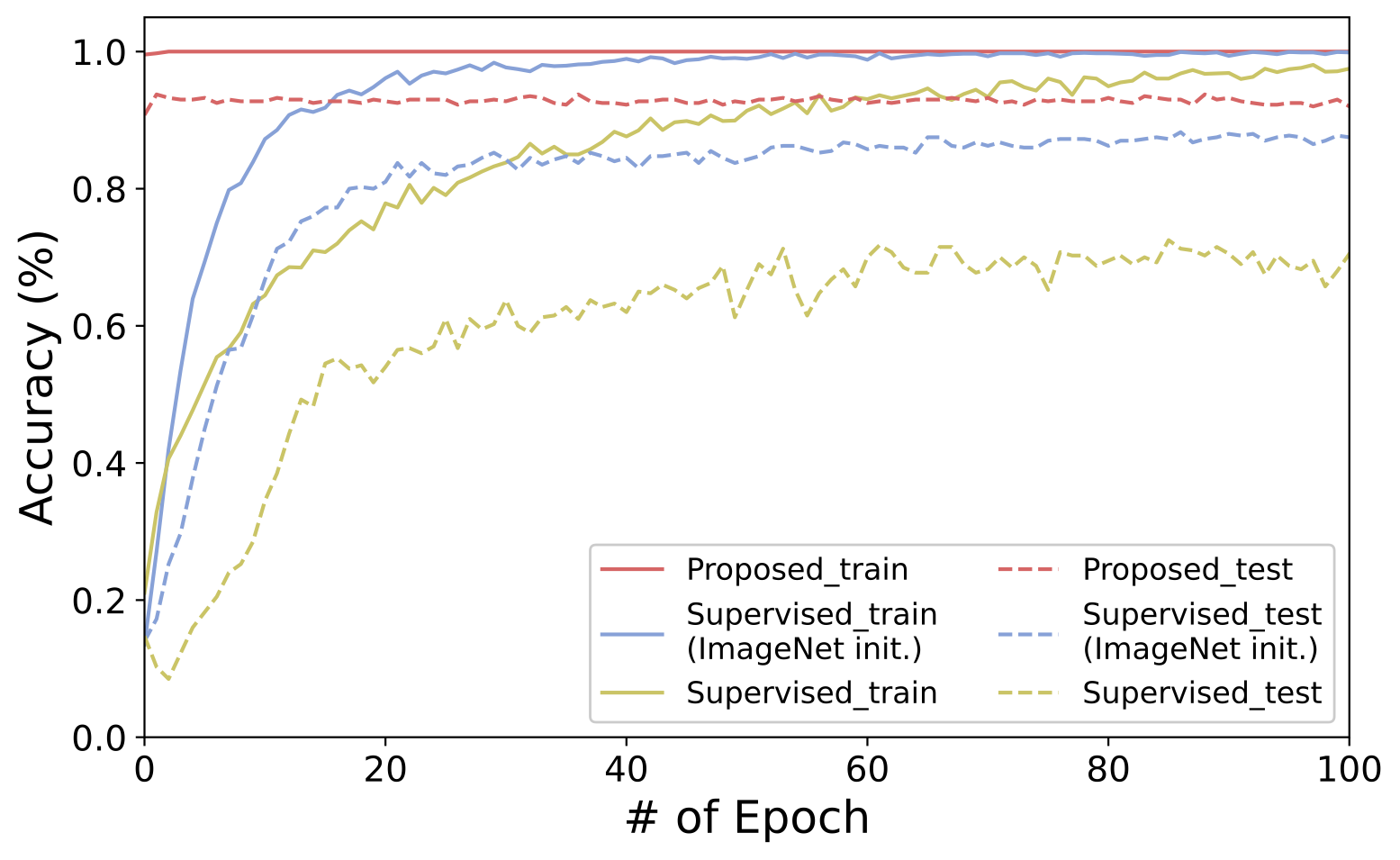}
    }
  \caption{The learning curve for each learning method under the condition of using 100\% label data of \textit{ESC10}. Each accuracy represents the average accuracy of each epoch in a 5-fold experiment.}
 \label{fig:generalization}
\end{figure}

We assume that the main factors in improving performance are stable training and generalization. 
The difference between \textit{last} and \textit{best} in the proposed method is always smaller than that of supervised, and smaller or similar to that of the weight initialized network. It implies that contrastive regularization covers the target task as we assumed and guides the network in a direction that helps generalization even when the data is scarce. These trends also appear in Fig \ref{fig:generalization}. 
The supervised training shows a typical learning curve. 
On the other hand, in the proposed method, the training accuracy increases rapidly, and both training and test accuracy remain constant. 
The rapid training seems to be due to contrastive regularization as well as warm-up training.
In addition to fast training set convergence, preventing overfitting also seems to be an effect of the proposed method. 
High training accuracy was quickly achieved through warm-up training, but test accuracy was also consistently high through contrastive regularization. Through this, it is expected that much effort to find the optimal point could be reduced.

\subsection{Ablation study}

% \begin{table}[]
% \centering
% \caption{\textit{ESC50} - ablation}
% \begin{tabular}{c|c}
% \hline
% \hline
% Proposed method      & 83.5±3.0 \\
% - Data mix      & 82.0±3.0 \\
% - External data & 79.5±1.6 \\
% Supervised      & 52.2±2.3 \\ \hline\hline
% \end{tabular}
% \end{table}

\begin{table}[ht]
\centering
\caption{Ablation study on \textit{ESC50} dataset.}
\label{table:ablation}
\begin{tabular}{cccc}
\hline 
Proposed & -Mixing & -Unlabeled  & Supervised\\ \hline
83.5 ± 3.0        & 82.0 ± 3.0      & 79.5 ± 1.6 &  73.6 ± 1.0      \\ \hline 
\end{tabular}
\end{table}

Table \ref{table:ablation} shows the results of excluding each element from the overall proposed method. 
In \texttt{-Unlabeled} condition, contrast regularization is added only within a labeled data set, with a performance improvement of 5.9\% points over the normal supervised condition. 
It shows that the proposed method is also helpful in a typical supervised setting. 
Without the data mixing strategy of the proposed method, \texttt{-Mixing}, which applies only time offset augmentation, there is a performance improvement of 8.4\% points over the supervised condition.
This implies that the increased diversity of tasks that external data can provide is a major factor in contrastive regularization. 

\section{Conclusion}
In this study, we proposed cross-domain semi-supervised training that works with completely different data distributions.
We introduced contrastive learning to the semi-supervised framework and proposed a novel augmentation method considering the audio characteristics. Experimental results have proven the effectiveness of the proposed method for audio event classification. 
Considering that the proposed method was implemented in a simplified form and the characteristics of the proposed method that it can be applied to any network, there is a lot of room for performance improvement.
In particular, we expect that various augmentation methods and large-batch, which are the main factors in contrastive learning, may improve the performance further.

% -------------------------------------------------------------------------
% Either list references using the bibliography style file IEEEtran.bst
\bibliographystyle{IEEEtran}
\bibliography{refs21}

\end{sloppy}
\end{document}